\documentclass{svmult}
\usepackage{graphicx}
\usepackage{amsmath,amssymb}

\newcommand{\be}{\begin{eqnarray}}
\newcommand{\ee}{\end{eqnarray}}

\begin{document}
\title*{Dark Energy from Brane-world Gravity }
\author{Roy Maartens}
\institute{Institute of Cosmology \& Gravitation, Portsmouth
University, Portsmouth~PO1~2EG, UK}

\maketitle

\begin{abstract}

Recent observations provide strong evidence that the universe is
accelerating. This confronts theory with a severe challenge.
Explanations of the acceleration within the framework of general
relativity are plagued by difficulties. General relativistic
models require a ``dark energy" field with effectively negative
pressure. An alternative to dark energy is that gravity itself may
behave differently from general relativity on the largest scales,
in such a way as to produce acceleration. The alternative approach
of modified gravity also faces severe difficulties, but does
provide a new angle on the problem. This review considers an
example of modified gravity, provided by brane-world models that
self-accelerate at late times. \footnote{Based on a talk at the
3rd Aegean Summer School, Chios, September 2005}

\end{abstract}

\section{Introduction}

\begin{figure}
\begin{center}
\includegraphics[height=3.5in,width=3.25in]{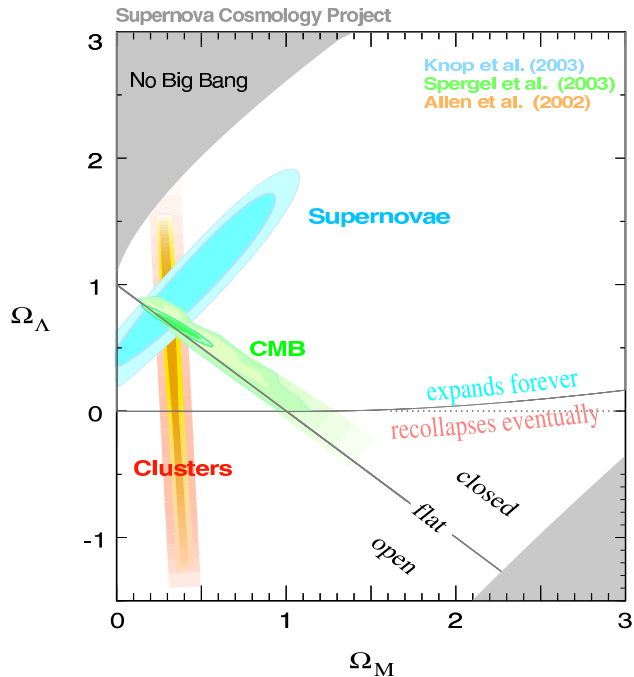}
\end{center}
\caption{Observational constraints in the
$(\Omega_\Lambda,\Omega_{M})$ plane (from~\cite{Knop:2003iy}).}
\label{sncmbgc}
\end{figure}

The current ``standard model" of cosmology -- the inflationary
cold dark matter model with cosmological constant (LCDM), based on
general relativity and particle physics (the minimal
supersymmetric extension of the Standard Model) -- provides an
excellent fit to the wealth of high-precision observational
data~\cite{Scott:2005uf}. In particular, independent data sets
from CMB anisotropies, galaxy surveys and SNe redshifts, provide a
consistent set of model parameters. For the fundamental energy
density parameters, this is shown in Fig.~\ref{sncmbgc}. The data
indicates that the cosmic energy budget is given by
\begin{equation}
\Omega_\Lambda\approx 0.7\,,~~ \Omega_{M}\approx 0.3\,,
\end{equation}
leading to the dramatic conclusion that the universe is undergoing
a late-time acceleration. The data further indicates that the
universe is (nearly) spatially flat, and that the primordial
perturbations are (nearly) scale-invariant, adiabatic and
Gaussian.

This standard model is remarkably successful, but we know that its
theoretical foundation, general relativity, breaks down at high
enough energies, usually taken to be at the Planck scale,
\begin{equation}
E \gtrsim M_p \sim 10^{16}\,\mbox{TeV}\,.
\end{equation}
The LCDM model can only provide limited insight into the very
early universe. Indeed, the crucial role played by inflation
belies the fact that inflation remains an effective theory without
yet a basis in fundamental theory. A quantum gravity theory will
be able to probe higher energies and earlier times, and should
provide a consistent basis for inflation, or an alternative that
replaces inflation within the standard cosmological model.

An even bigger theoretical problem than inflation is that of the
recent accelerated expansion of the universe. Within the framework
of general relativity, the acceleration must originate from a dark
energy field with effectively negative pressure ($w \equiv
p/\rho<-{1\over3}$), such as vacuum energy ($w=-1$) or a
slow-rolling scalar field (``quintessence", $w>-1$). So far, none
of the available models has a natural explanation.

For the simplest option of vacuum energy, i.e., the LCDM model,
the incredibly small value of the cosmological constant
\begin{eqnarray}
&& \rho_{\Lambda,\text{obs}}={\Lambda \over 8\pi G}\sim H_0^2M_P^2
\sim (10^{-33}\,\mbox{eV})^2(10^{19}\,\mbox {GeV})^2=10^{-57}\,
\mbox{GeV}^4, \\ && \rho_{\Lambda,\text{theory}}\sim
 M_{\text{fundamental}}^4 > 1~\mbox{TeV}^4 \gg
\rho_{\Lambda,\text{obs}}\,,
\end{eqnarray}
cannot be explained by current particle physics. In addition, the
value needs to be incredibly fine-tuned,
\begin{equation}
\Omega_{\Lambda}\sim \Omega_M\,,
\end{equation}
which also has no natural explanation. Quintessence models attempt
to address the fine-tuning problem, but do not succeed fully --
and also cannot address the problem of how $\Lambda$ is set
exactly to 0. Quantum gravity will hopefully provide a solution to
the problems of vacuum energy and fine-tuning.

Alternatively, it is possible that there is no dark energy, but
instead a low-energy/ large-scale (i.e., ``infrared") modification
to general relativity that accounts for late-time acceleration.
Schematically, we are modifying the geometric side of the field
equations,
 \be
G_{\mu\nu}+G^{\rm dark}_{\mu\nu} = 8\pi G T_{\mu\nu}\,,
\label{mod}
 \ee
rather than the matter side,
 \be
G_{\mu\nu} = 8\pi G \left(T_{\mu\nu}+ T^{\rm dark}_{\mu\nu}
\right)\,,
 \ee
as in general relativity.

It is important to stress that a consistent modification of
general relativity requires a covariant formulation of the field
equations in the general case, i.e., including inhomogeneities and
anisotropies. It is not sufficient to propose ad hoc modifications
of the Friedman equation, of the form
 \be
f(H^2) = {8\pi G\over 3} \rho ~~\mbox{or}~ ~ H^2  = {8\pi G\over
3} g(\rho) \,,
 \ee
for some functions $f$ or $g$. We can compute the SNe redshifts
using this equation -- but we {\em cannot} compute the density
perturbations without knowing the covariant parent theory that
leads to such a modified Friedman equation.

\begin{figure}[!bth]\label{brane}
\begin{center}
\includegraphics[height=3in,width=4in]{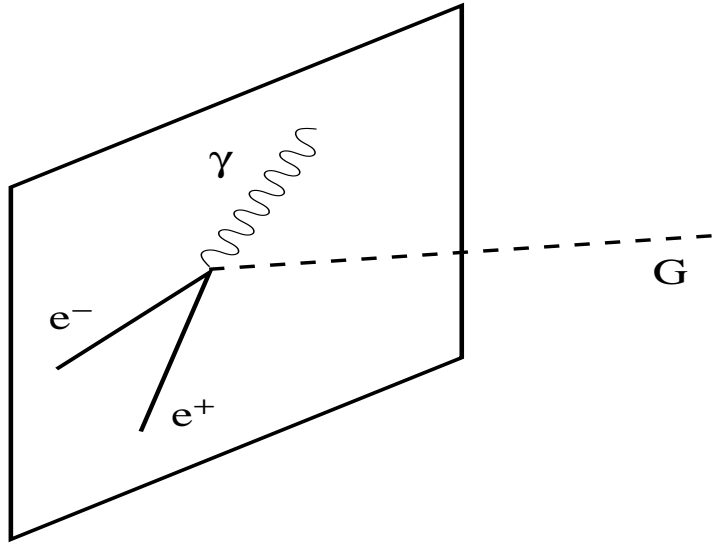}
\caption{The confinement of matter to the brane, while gravity
propagates in the bulk (from~\cite{Cavaglia:2002si}).}
\end{center}
\end{figure}

An infra-red modification to general relativity could emerge
within the framework of quantum gravity, in addition to the
ultraviolet modification that must arise at high energies in the
very early universe. The leading candidate for a quantum gravity
theory, string theory, is able to remove the infinities of quantum
field theory and unify the fundamental interactions, including
gravity. But there is a price -- the theory is only consistent in
9 space dimensions. Branes are extended objects of higher
dimension than strings, and play a fundamental role in the theory,
especially D-branes, on which open strings can end. Roughly
speaking, open strings, which describe the non-gravitational
sector, are attached at their endpoints to branes, while the
closed strings of the gravitational sector can move freely in the
higher-dimensional ``bulk" spacetime. Classically, this is
realised via the localization of matter and radiation fields on
the brane, with gravity propagating in the bulk (see
Fig.~\ref{brane}).

The implementation of string theory in cosmology is extremely
difficult, given the complexity of the theory. This motivates the
development of phenomenology, as an intermediary between
observations and fundamental theory. (Indeed, the development of
inflationary cosmology has been a very valuable exercise in
phenomenology.) Brane-world cosmological models inherit key
aspects of string theory, but do not attempt to impose the full
machinery of the theory. Instead, drastic simplifications are
introduced in order to be able to construct cosmological models
that can be used to compute observational predictions
(see~\cite{Maartens:2003tw} for reviews in this spirit).
Cosmological data can then be used to constrain the brane-world
models, and hopefully thus provide constraints on string theory,
as well as pointers for the further development of string theory.

It turns out that even the simplest brane-world models are
remarkably rich -- and the computation of their cosmological
perturbations is remarkably complicated, and still incomplete.
Here I will describe brane-world cosmologies of
Dvali-Gabadadze-Porrati (DGP) type~\cite{Dvali:2000rv}. These are
5-dimensional models, with an infinite extra dimension. (We
effectively assume that 5 of the extra dimensions in the ``parent"
string theory may be ignored at low energies.)

\section{KK modes of the graviton}

The brane-world mechanism, whereby matter is confined to the brane
while gravity accesses the bulk, means that extra dimensions can
be much larger than in the conventional Kaluza-Klein (KK)
mechanism, where matter and gravity both access all dimensions.
The dilution of gravity via the bulk effectively weakens gravity
on the brane, so that the true, higher-dimensional Planck scale
can be significantly lower than the effective 4D Planck scale
$M_p$.

The higher-dimensional graviton has massive 4D modes felt on the
brane, known as KK modes, in addition to the massless mode of 4D
gravity. From a geometric viewpoint, the KK modes can also be
understood via the fact that the projection of the null graviton
5-momentum $p^{(5)}_a$ onto the brane is timelike. If the unit
normal to the brane is $n_a$, then the induced metric on the brane
is
 \be
g_{ab}= g^{(5)}_{ab}-n_an_b\,,~  g^{(5)}_{ab}n^an^b=1\,,~
g_{ab}n^b=0 \,,
 \ee
and the 5-momentum may be decomposed as
 \be
p^{(5)}_a=mn_a+ p_a\,,~ p_an^a=0\,,~m=p^{(5)}_a\, n^a \,,
 \ee
where $p^a=g^{ab}p^{(5)}_b$ is the projection along the brane,
depending on the orientation of the 5-momentum relative to the
brane. The effective 4-momentum of the 5D graviton is thus $p_a$.
Expanding $g^{(5)}_{ab}p_{(5)}^a p_{(5)}^b=0$, we find that
 \be
g_{\mu\nu}p^\mu p^\nu =-m^2\,,
 \ee
using coordinates $x^a=(x^\mu,y)$, where $y$ is along the extra
dimension. It follows that the 5D graviton has an effective mass
$m$ on the brane. The usual 4D graviton corresponds to the zero
mode, $m=0$, when $p^{(5)}_a$ is tangent to the brane.

The extra dimensions lead to new scalar and vector degrees of
freedom on the brane. The spin-2 5D graviton is represented by a
metric perturbation $h^{(5)}_{ab}$ that is transverse traceless:
 \be
g^{(5)}_{ab} \to g^{(5)}_{ab}+ h^{(5)}_{ab}\,,~
h^{(5)a}{}_{a}=0=\nabla^{(5)}_b\, h^{(5)b}{}_{a}\,.
 \ee
In a suitable gauge, $ h^{(5)}_{ab}$ contains a 3D transverse
traceless perturbation $h_{ij}$ (where $x^\mu=(x^0,x^i)$), a 3D
transverse vector perturbation $\Sigma_i$ and a scalar
perturbation $\beta$, which each satisfy the 5D wave equation:
 \be \label{5dg}
&& h^i{}_i=0=\nabla_j h^{ij}\,,~ \nabla_i \Sigma^i=0\,,\\
 &&(\nabla_\mu \nabla^\mu +\partial_y^2)\left(
 \begin{array}{c} \beta\\ \Sigma_i \\
h_{ij} \end{array} \right)=0\,.
 \ee
The 5 degrees of freedom (polarizations) in the 5D spin-2 graviton
are felt on the brane as:
\begin{itemize}
\item a 4D spin-2 graviton $h_{ij}$ (2 polarizations) \item a 4D
spin-1 gravi-vector (gravi-photon) $\Sigma_i$ (2 polarizations)
\item a 4D spin-0 gravi-scalar $\beta$.
\end{itemize}
The massive modes of the 5D graviton are represented via massive
modes in all 3 of these fields on the brane. The standard 4D
graviton corresponds to the massless zero-mode of $h_{ij}$.

\section{DGP type brane-worlds: self-accelerating cosmologies}

Could the late-time acceleration of the universe be a
gravitational effect?\footnote{Note that this would not remove the
problem of explaining why the vacuum energy does not gravitate.}
An historical precedent is provided by attempts to explain the
anomalous precession of Mercury's perihelion by a ``dark planet".
In the end, it was discovered that a modification to Newtonian
gravity was needed.

An alternative to dark energy plus general relativity is provided
by models where the acceleration is due to modifications of
gravity on very large scales, $r\gtrsim H_0^{-1}$. It is very
difficult to produce infrared corrections to general relativity by
modifying the 4D Einstein-Hilbert action,
 \be
\int d^4x\,\sqrt{-g}\,R ~\to~ \int d^4x\,\sqrt{-g}\,f(R,
R_{\mu\nu}R^{\mu\nu},\dots)\,.
 \ee
Typically, instabilities arise or the action has no natural
motivation. The DGP brane-world offers a higher-dimensional
approach to the problem, which effectively has infinite extra
degrees of freedom from a 4D viewpoint.

Most brane-world models modify general relativity at high
energies. The main examples are those of Randall-Sundrum (RS)
type~\cite{Randall:1999vf}, where a Friedman-Robertson-Walker
brane is embedded in an anti de Sitter bulk, with curvature radius
$\ell$. At low energies $H\ell \ll 1$, the zero-mode of the
graviton dominates on the brane, and general relativity is
recovered to a good approximation. At high energies, $H\ell\gg 1$,
the massive modes of the graviton dominate over the zero mode, and
gravity on the brane behaves increasingly in a 5D way. On the
brane, the standard conservation equation holds,
 \be
\dot\rho+3H(\rho+p)=0\,,
  \ee
but the Friedmann equation is modified by an ultraviolet
correction:
 \be\label{mf}
H^2 = \frac{8\pi G}{3} \rho\left(1+{2\pi G \ell^2\over 3}\rho
\right) + \frac{\Lambda}{3} \,.
 \ee
The $\rho^2$ term is the ultraviolet term. At low energies, this
term is negligible, and we recover $H^2 \propto \rho+\Lambda/8\pi
G $. At high energies, gravity ``leaks" off the brane and
$H^2\propto \rho^2$. This 5D behaviour means that a given energy
density produces a greater rate of expansion than it would in
general relativity. As a consequence, inflation in the early
universe is modified in interesting ways~\cite{Maartens:2003tw}.

In the DGP case the bulk is 5D Minkowski spacetime. Unlike the AdS
bulk of the RS model, the Minkowski bulk has infinite volume.
Consequently, there is no normalizable zero-mode of the graviton
in the DGP brane-world. Gravity leaks off the 4D brane into the
bulk at large scales. At small scales, gravity is effectively
bound to the brane and 4D dynamics is recovered to a good
approximation. The transition from 4- to 5D behaviour is governed
by a crossover scale $r_c$; the weak-field gravitational potential
behaves as
\begin{equation}
\Psi \sim \left\{ \begin{array}{lll} r^{-1} & \mbox{for} & r\ll
r_c
\\ r^{-2} & \mbox{for} & r\gg r_c \end{array}\right.
\end{equation}
Gravity leakage at late times initiates acceleration -- not due to
any negative pressure field, but due to the weakening of gravity
on the brane. 4D gravity is recovered at high energy via the
lightest KK modes of the graviton, effectively via an ultralight
metastable graviton.

\begin{figure}[!bth]\label{fig2}
\begin{center}
\includegraphics[height=3.5in,width=4.5in]{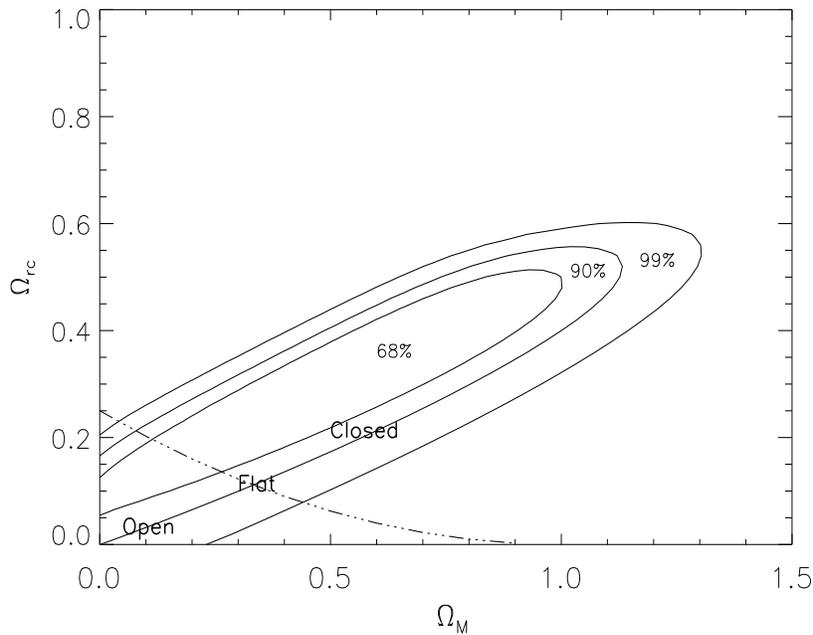}
\caption{Constraints from SNe redshifts on DGP models.
(From~\cite{Deffayet:2002sp}.)}
\end{center}
\end{figure}

The energy conservation equation remains the same as in general
relativity, but the Friedman equation is modified:
\begin{eqnarray}
&& \dot\rho+3H(\rho+p)=0\,,\label{ec} \\ && H^2-{H \over r_c}=
{8\pi G \over 3}\rho\,. \label{f}
\end{eqnarray}
This shows that at early times, $Hr_c \gg 1$, the general
relativistic Friedman equation is recovered. By contrast, at late
times in a CDM universe, with $\rho\propto a^{-3}\to0$, we have
\begin{equation}
H\to H_\infty= {1\over r_c}\,.
\end{equation}
Since $H_0>H_\infty$, in order to achieve self-acceleration at
late times, we require
 \be
r_c\gtrsim H_0^{-1}\,,
 \ee
and this is confirmed by fitting SNe observations, as shown in
Fig.~\ref{fig2}. This comparison is aided by introducing a
dimensionless cross-over parameter,
 \be
\Omega_{r_c}={1\over 4(H_0r_c)^2}\,,
 \ee
and the LCDM relation,
 \be
\Omega_M+\Omega_\Lambda+\Omega_K=1\,,
 \ee
is modified to
 \be
\Omega_M+ 2\sqrt{\Omega_{r_c}}\sqrt{1-\Omega_K}+\Omega_K=1\,.
 \ee

It should be emphasized that the DGP Friedman equation~(\ref{f})
is derived covariantly from a 5D gravitational action,
 \be
\int_{\rm bulk} d^5x\,\sqrt{-g^{(5)}}\,R^{(5)}+ r_c  \int_{\rm
brane} d^4x\,\sqrt{-g}\,R \,.
  \ee

LCDM and DGP can both account for the SNe observations, with the
fine-tuned values $\Lambda\sim H_0^2$ and $r_c\sim H_0^{-1}$
respectively. This degeneracy may be broken by observations based
on structure formation, since the two models suppress the growth
of density perturbations in different
ways~\cite{Lue:2004rj,wrong}. The distance-based SNe observations
draw only upon the background 4D Friedman equation~(\ref{f}) in
DGP models -- and therefore there are quintessence models in
general relativity that can produce precisely the same SNe
redshifts as DGP~\cite{Linder:2005in}. By contrast, structure
formation observations require the 5D perturbations in DGP, and
one cannot find equivalent general relativity
models~\cite{Koyama:2005kd}.

For LCDM, the analysis of density perturbations is well
understood. For DGP it is much more subtle and complicated.
Although matter is confined to the 4D brane, gravity is
fundamentally 5D, and the bulk gravitational field responds to and
backreacts on density perturbations. The evolution of density
perturbations requires an analysis based on the 5D nature of
gravity. In particular, the 5D gravitational field produces an
anisotropic stress on the 4D universe. Some previous results are
based on inappropriately neglecting this stress and all 5D effects
-- as a consequence, the 4D Bianchi identity on the brane is
violated, i.e., $\nabla^\nu G_{\mu\nu} \neq 0$, and the results
are inconsistent.

\begin{figure}[t]
\centerline{
\includegraphics[width=9cm]{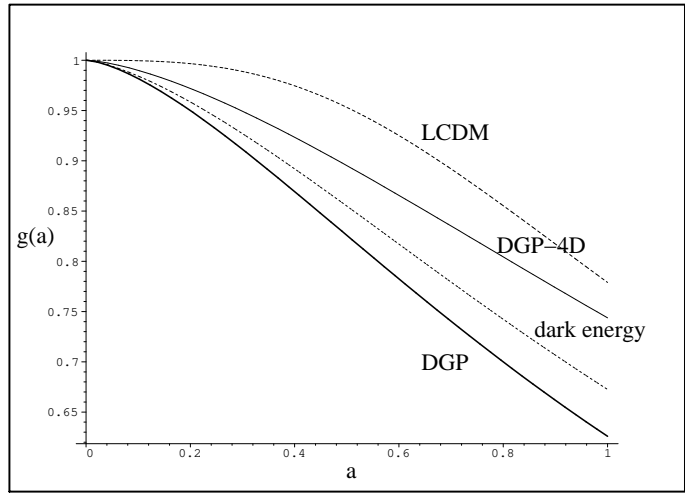}}
\caption{The growth factor $g(a)=\Delta(a)/a$ for LCDM (long
dashed) and DGP (solid, thick), as well as for a dark energy model
with the same expansion history as DGP (solid, thick). DGP-4D
(solid, thin) shows the incorrect result in which the 5D effects
are set to zero. (From~\cite{Koyama:2005kd}.)} \label{fig:fig1}
\end{figure}

When the 5D effects are incorporated~\cite{Koyama:2005kd}, the 4D
Bianchi identity is satisfied. (The results
of~\cite{Koyama:2005kd} confirm and generalize those
of~\cite{Lue:2004rj}.) The consistent modified evolution equation
for density perturbations is
\begin{equation}\label{dpe}
\ddot{\Delta} + 2 H \dot{\Delta}=4\pi G \left\{1 -
\frac{(2Hr_c-1)}{3[2(Hr_c)^2-2 Hr_c+1]} \right\} \rho \Delta\,,
\end{equation}
where the term in braces encodes the 5D correction. The linear
growth factor, $g(a)=\Delta(a)/a$ (i.e., normalized to the flat
CDM case, $\Delta \propto a$), is shown in Fig.~\ref{fig:fig1}.

It must be emphasized that these results apply on subhorizon
scales. On superhorizon scales, where the 5D effects are
strongest, the problem has yet to be solved. This solution is
necessary before one can compute the large-angle CMB anisotropies
-- any prediction of the large-scale anisotropies without solving
the 5D perturbation problem is unreliable.

It should also be remarked that the late-time asymptotic de Sitter
solution in DGP cosmological models has a ghost
problem~\cite{Gorbunov:2005zk}, which makes the quantum vacuum
unstable and which may have implications for the analysis of
density perturbations. As a classical model, the DGP is covariant
and consistent, and we effectively assume that the ghost problem
will be solved by a quantum gravity ultraviolet completion of the
model.

\section{Conclusion}

In conclusion, DGP brane-world models, which are inspired by ideas
from string theory, provide a rich and interesting phenomenology
for modified gravity. These models can account for the late-time
acceleration without the need for dark energy -- gravity leakage
from the 4D brane at large scales leads to self-acceleration. The
5D graviton, i.e., its KK modes, plays a crucial role, which has
been emphasized in this article.

\[ \]{\bf Acknowledgenments:}

I thank the organisers for the invitation to present this work,
which was supported by PPARC.



\begin{thebibliography}{99}

\bibitem{Scott:2005uf}
See, e.g.,  D.~Scott,
  arXiv:astro-ph/0510731.

\bibitem{Knop:2003iy}
  R.~A.~Knop {\it et al.}  [The Supernova Cosmology Project Collaboration],
  Astrophys.\ J.\  {\bf 598}, 102 (2003)
  [arXiv:astro-ph/0309368].

\bibitem{Cavaglia:2002si}
  M.~Cavaglia,
  Int.\ J.\ Mod.\ Phys.\ A {\bf 18}, 1843 (2003)
  [arXiv:hep-ph/0210296].

\bibitem{Maartens:2003tw}
  R.~Maartens,
  Living Rev.\ Rel.\  {\bf 7}, 7 (2004)
  [arXiv:gr-qc/0312059];\\
  P.~Brax, C.~van de Bruck and A.~C.~Davis,
  Rept.\ Prog.\ Phys.\  {\bf 67}, 2183 (2004)
  [arXiv:hep-th/0404011];\\
  V.~Sahni,
  arXiv:astro-ph/0502032;\\
  R.~Durrer,
  AIP Conf.\ Proc.\  {\bf 782}, 202 (2005)
  [arXiv:hep-th/0507006];\\
  D.~Langlois,
  arXiv:hep-th/0509231;\\
  A.~Lue,
  Phys.\ Rept.\  {\bf 423}, 1 (2006)
  [arXiv:astro-ph/0510068];\\
  D.~Wands,
  arXiv:gr-qc/0601078.

\bibitem{Dvali:2000rv}
  G.~R.~Dvali, G.~Gabadadze and M.~Porrati,
  Phys.\ Lett.\ B {\bf 484}, 112 (2000)
  [arXiv:hep-th/0002190];\\
  C.~Deffayet,
  Phys.\ Lett.\ B {\bf 502}, 199 (2001)
  [arXiv:hep-th/0010186].

\bibitem{Randall:1999vf}
  L.~Randall and R.~Sundrum,
  Phys.\ Rev.\ Lett.\  {\bf 83}, 4690 (1999)
  [arXiv:hep-th/9906064];\\
  P.~Binetruy, C.~Deffayet, U.~Ellwanger and D.~Langlois,
  Phys.\ Lett.\ B {\bf 477}, 285 (2000)
  [arXiv:hep-th/9910219].

\bibitem{Deffayet:2002sp}
  C.~Deffayet, S.~J.~Landau, J.~Raux, M.~Zaldarriaga and P.~Astier,
  Phys.\ Rev.\ D {\bf 66}, 024019 (2002)
  [arXiv:astro-ph/0201164].

\bibitem{Lue:2004rj}
  A.~Lue, R.~Scoccimarro and G.~D.~Starkman,
  Phys.\ Rev.\ D {\bf 69}, 124015 (2004)
  [arXiv:astro-ph/0401515];\\
  A.~Lue and G.~Starkman,
  Phys.\ Rev.\ D {\bf 67}, 064002 (2003)
  [arXiv:astro-ph/0212083].

\bibitem{wrong}
  Y.~S.~Song,
  Phys.\ Rev.\ D {\bf 71}, 024026 (2005)
  [arXiv:astro-ph/0407489];\\
  L.~Knox, Y.~S.~Song and J.~A.~Tyson,
  arXiv:astro-ph/0503644;\\
  M.~Ishak, A.~Upadhye and D.~N.~Spergel,
  arXiv:astro-ph/0507184;\\
  I.~Sawicki and S.~M.~Carroll,
  arXiv:astro-ph/0510364.

\bibitem{Linder:2005in}
  E.~V.~Linder,
  Phys.\ Rev.\ D {\bf 72}, 043529 (2005)
  [arXiv:astro-ph/0507263].

\bibitem{Koyama:2005kd}
  K.~Koyama and R.~Maartens,
  JCAP {\bf 0610}, 016 (2006)
  [arXiv:astro-ph/0511634].

\bibitem{Gorbunov:2005zk}
  D.~Gorbunov, K.~Koyama and S.~Sibiryakov,
  Phys.\ Rev.\ D, to appear
  [arXiv:hep-th/0512097].


\end{thebibliography}
\end{document}